\documentstyle[epsfig]{mn}

\title[Lens modelling and $H_{0}$ estimate in quadruply lensed systems]{Lens modelling and $H_{0}$ estimate in quadruply lensed systems}

\author[C. Tortora et al.]{C. Tortora$^{1}$ \thanks{Corresponding author\,: {\tt ctortora@na.infn.it}},
 E. Piedipalumbo$^{1}$,
 V.F. Cardone$^{2}$ \\
$^1$ Dipartimento di Scienze fisiche, Universit\`{a} degli studi di Napoli
``Federico II'', and INFN, Sezione di Napoli, \\ Complesso Universitario di
Monte S. Angelo, Via Cinthia, Edificio N, 80126 Napoli, Italy\\ $^2$
Dipartimento di Fisica ``E.R. Caianiello'', Universit{\`{a}} di Salerno and
INFN, Sezione di Napoli, \\ Gruppo Collegato di Salerno, Via S. Allende,
84081 - Baronissi (Salerno), Italy}

\date{Accepted xxx.  Received yyy. in original form}

\begin{document}
\maketitle

\begin{abstract}
We present a numerical method to estimate the lensing parameters
and the Hubble constant $H_0$ from quadruply imaged gravitational
lens systems. The lens galaxy is modeled using both separable
deflection potentials and constant mass\,-\,to\,-\,light ratio
profiles, while possible external perturbations have been taken
into account introducing an external shear. The model parameters
are recovered inverting the lens and the time delay ratio
equations and imposing a set of physically motivated selection
criteria. We investigate correlations among the model parameters
and the Hubble constant. Finally, we apply the codes to the real
lensed quasars PG 1115+080 and RX J0911+0551, and combine the
results from these two systems to get $H_0 = 56 {\pm} 23 \ {\rm km
\ s^{-1} \ Mpc^{-1}}$. In addition, we are able to fit to the
single systems a general elliptical potential with a non fixed
angular part, and then we model the two lens systems with the same
potential and a shared $H_{0}$: in this last case we estimate
$H_{0}=49_{-11}^{+6} \ \textrm{Km} \ \textrm{s}^{-1}\
\textrm{Mpc}^{-1}$.
\end{abstract}


\section{Introduction}
Gravitational lensing is one of the  main tools to obtain
information about the structure and evolution of the universe. In
particular, time delay measurements are a recent primary distances
indicator, furnishing a new method to estimate the Hubble constant
$H_0$, which determines the present expansion rate of the universe
(see, for instance, \cite{Na-Ba98,Ko-Sc03}).

Actually, in 1964, Refsdal proposed to estimate $H_{0}$
\cite{Ref64a,Ref64b,Ref66} from multiply imaged QSOs by the
measurements of the delays in the arrive time between  light rays
coming from the different images, which follow different optical
paths. It is not difficult to show that the time delay between two
images due to a gravitational lens can be factorized in two
pieces: the first one depends on cosmological parameters and is
inversely proportional to $H_0$, while the second one is
determined by the lens model only. Thus, having measured time
delays among images of a lensed QSO, once we fix the cosmological
parameters, we can obtain a direct estimate of $H_{0}$ provided
that the lens model has been recovered from the lensing
constraints, or is known in an other independent way. Nowadays,
there are more than sixty multiple image systems \cite{castles},
but only about ten of them have measured time delays. However,
this number is increasing day by day, and in the future it will be
possible to measure time delays for many other systems.

It is well known hat the most significant uncertainty affecting
the estimate of $H_0$ with the Refsdal method is only related to
the mass model used. In the usual approach the model parameters
are recovered by fitting some  parametric models to the available
constraints through $\chi^{2}$ minimization techniques. Instead,
other authors (see \cite{Wi-Saha2000}) carried out the
``pixellated lens'' modelling, that describes the mass
distribution by a large number of discrete pixels with arbitrary
densities, so determining the Hubble constant by means of a set of
physical motivated constraints. A compromise between these two
approaches consists in the numerical solution of a set of non
linear equations:  in a previous paper (\cite{HERQULES}, hereafter
CCRP02), for instance, some of us applied a semianalytical
technique to fit the separable potential of the form $\psi =
r^{\alpha} F(\theta)$ and were able to develop an algorithm to
estimate $H_0$ without the need to give an explicit expression for
the shape function $F(\theta)$.

Here, we further improve the general method in \cite{PULP} (hereafter
CCRP01) and CCRP02, implementing a set of {\it Mathematica} codes to shape
quadruple lens systems and obtain a better estimate of the Hubble constant.
Such a method in fact allows us to manage a still wider class of lens
models and to obtain useful information about lens parameters and $H_{0}$,
eliminating possible not physical solutions.

Actually, we consider both separable models, specifying the angular part
$F(\theta)$, and models with constant mass-to-light ratio. As usually in
literature, we develop the potential of an external object contributing to
the lensing phenomenon to second order, and hence its effect on the total
deflection potential translates into an external shear term \cite{SEF}. We
constrain the lens models using the image positions with respect to the
lens centre and the time delay ratios, which allow us to highly constrain
the used models, even if the equations to be solved become more complex.
Once the results from all the possible models are collected together, such
a procedure offers the advantage of giving an estimate of the Hubble
constant which is in a sense {\it marginalized} with respect to the lens
models.

Moreover, our numerical method allows to pursue a new kind of
approach to the lens modelling, that considers at the same time
several lens systems. Actually, following \cite{SW2004}, we try to
model two lensed systems with the elliptical potential
$\psi=r^{\alpha} F(\theta)$ and an external shear, based on a
strong hypothesis we make \emph{ab initio}: we suppose that the
lens systems has a shared $H_{0}$. In this way we are able to
create a two-system model that can give a more complete estimate
of $H_{0}$.

The outline of the paper is as follows. In Sect. \ref{Lens equation and
time delay} we write the lens equations and the time delay function in
polar coordinates. A careful presentation of the models we use is given in
Sect. \ref{Lens models}, while Sect. \ref{Single system} is devoted to the
presentation of the numerical method used to `fit' single lens models to
systems. We also discuss the constraints used to select physically
motivated solutions and the statistical approach to obtain the final
estimate of the parameters. Then in Sect. \ref{two-lens-model} we present a
new approach to shape two lens systems with a shared $H_{0}$. In order to
verify whether our codes are able or not to recover the correct values of
parameters, we build simulated systems to which we apply the codes as
described in Sect. \ref{Simulated systems}, where, in addition, we analyze
the biases and the uncertainties in the use of a model respect to another
one. In Sect. \ref{correlations}, we use the simulated systems to
investigate the existence of degeneracies among the model parameters,
recovering some interesting scaling laws. Sect. \ref{applications} is
dedicated to the application of our procedure to two real quadruple
systems, PG 1115+080 and RX J0911+0551: we obtain an estimate of the
lensing parameters and the Hubble constant, also taking into account the
contribution of changing the cosmological parameters to the uncertainty on
$H_0$. Finally, we present a discussion of our results and future
improvements in Sect. \ref{conclusion}.

\section{Lens equation and time delay}\label{Lens equation and time delay}

Let us choose a rectangular system $(x,y)$ centered on the lens
galaxy and with axes pointing towards West and North,
respectively. Let $(r,\theta)$ be the polar coordinates, being
$\theta$ the position angle measured from North to East, connected
to the rectangular ones through the following coordinate
transformation:

\begin{equation}
x = - r \sin \theta , \hspace{0.5 cm}  y = r \cos \theta \ .
\end{equation}
Time delay function, i.e. time delay of a generic path from source
to observer, is given by \cite{BlNar,ZP2001}:
\begin{eqnarray}
\lefteqn{ \Delta t=h^{-1} \tau_{100}} \nonumber \\
 & \bigg[\frac{1}{2} r^{2} -r r_{s} \cos (\theta - \theta_{s}) +\frac{1}{2}r_{s}^{2} - \psi (r, \theta)\bigg] ,\label{timedelay}
\end{eqnarray}
where $(r,\theta)$ determines the impact position of the generic path
on the lens plane (with $r$ measured in arcsec), $(r_{s},\theta_{s})$
is the unknown source position, and $\psi(r,\theta)$ is the
adimensional lensing potential. Then, $h$ is the Hubble constant
$H_{0}$ in units of $100 \ {\rm km \ s^{-1} \ Mpc^{-1}}$, while
$\tau_{100}$ is a typical time delay linked to cosmological
parameters,  and defined as:
\begin{equation}
\tau _{100} \equiv \frac{D_{d}D_{s}}{D_{ds}} \frac{1+z_{d}}{c},
\end{equation}
where $D_{d}$, $D_{s}$ and $D_{ds}$ are  observer\,-\,lens,
observer\,-\,source,  and lens\,-\,source angular diameter distances,
calculated with $H_{0}=100 \ {\rm km \ s^{-1} \ Mpc^{-1}}$, and $z_{d}$ is
the lens redshift. This factor contains all the cosmological information,
since the distance depend on the other cosmological parameters: i.e. the
density matter parameter $\Omega_{m}$, the ``dark energy'' contribution
$\Omega_{X}$ (see \cite{Peeb02,cal+al98,Dem03}) and the smoothness
parameters $\tilde{\alpha}$ introduced in \cite{D-R1,D-R2,D-R3} to take
into account the inhomogeneities of the universe.  If $i$ and $j$ are two
images, the time delay between them is $\Delta t_{ij}=\Delta t_{i}-\Delta
t_{j}$.

According to Fermat principle, the images lie at the critical points of $\Delta t$, so one can obtain lens equations minimizing it.
 We get:

\begin{equation}
r-r_{s} \cos (\theta - \theta_{s})=\frac{\partial \psi}{\partial r},\label{eqlens1}
\end{equation}
\begin{equation}
r_{s} \sin (\theta - \theta_{s}) =\frac{1}{r}\frac{\partial \psi}{\partial \theta}.\label{eqlens2}
\end{equation}
We shall use a lensing potential formed by the sum of two terms:

\begin{equation}
\psi(r, \theta)= \psi_{lens}(r, \theta)+\psi_{ext}(r,\theta),
\end{equation}
where $\psi_{lens}(r, \theta)$ is the contribution of the lens galaxy,
and $\psi_{ext}(r,\theta)$ is the external perturbation. The first
term is connected with the mass distribution of the galaxy through the
Poisson equation:

\begin{equation}
\nabla^2 \psi_{lens}(r,\theta) = 2 \kappa (r, \theta),
\end{equation}
being $\kappa(r, \theta)$ the convergence, i.e., the adimensional
surface mass density, defined as:

\begin{equation}
\kappa(r, \theta)  \equiv \frac{\Sigma (r, \theta)}{\Sigma_{crit}},
\end{equation}
where $\Sigma_{crit}=\frac{c^{2}D_{s}}{4 \pi G D_{d}D_{ds}}$.
The second term describes the effects on the lensing system of the
environment, i.e., of the cluster of galaxies which the lens galaxy
belongs to, or an external group of galaxies. We describe this
contribution developing the lensing potential of the external
perturbation to the  second order:

\begin{equation}
\psi_{ext}(r, \theta) = \psi_{shear}(r,\theta)=-\frac{1}{2} \gamma r^{2} \cos2(\theta - \theta_{\gamma}),
\end{equation}
where $\gamma$ is the external shear and $\theta_{\gamma}$ is the shear angle, oriented from North to East and pointing
towards the external perturbation.
\section{Lens models}\label{Lens models}
The majority of the lens galaxies are early-type galaxies, since
lensing selects galaxies by mass and the fact that early type
galaxies are more massive than spirals overwhelms the fact that
spirals are slightly more numerous. Early type galaxies show a
wide variety of optical shapes including oblate, prolate and
triaxial spheroids. Moreover, also their mass distribution is not
yet fully understood: stellar dynamics and X ray observations all
suggest that the early type galaxies are dominated by dark matter
halos; on the other hand, some dynamical studies recently
performed using planetary nebulae as tracers in the halos of
early-type galaxies show evidence of a universal declining
velocity dispersion profile, and dynamical models indicate the
presence of little dark matter within $5 R_{e}$ -- implying halos
either not as massive or not as centrally concentrated as CDM
predicts \cite{rom1}. So not only a fundamental hypothesis of the
CDM paradigm have been remained up to now largely unverified --
that there should be similarly extended, massive dark halos around
ellipticals --, but predictions about the detailed halo properties
have not been testable. Gravitational lensing is actually an
unique tool to studying elliptical halos, and face the question
about the nature and the distribution of dark matter in early type
galaxies, with implications on the estimation of the Hubble
constant. In order to consider both a luminous-dominated and
dark-dominated component in the main lens galaxy, we consider two
different classes of models\,: separable potentials and constant
mass\,-\,to\,-\,light ($M/L$) mass profiles.

\begin{table}
\begin{center}
\caption{Separable deflection potentials.}\label{tabella-potenziali-separabili}
\vspace{0.5cm}
\begin{tabular}{|@{}c@{}|@{}c@{}|}
\hline
\rule{0pt}{2.7ex} Model  &  $\psi(r, \theta)$\\
\hline \hline
\rule{0pt}{2.7ex}{} Model 1{}  & $\frac{b^{2-\alpha}}{\alpha} r^{\alpha}$\\
\hline
\rule{0pt}{2.7ex}{} Model 2 {} & $b  \sqrt{\sin (\theta - \theta_{q})^{2}+q^{-2}\cos (\theta - \theta_{q})^{2}} \hspace{0.1cm} r$ {}\\
\hline
\rule{0pt}{2.7ex} {} Model 3 {} & {} $\frac{b^{2-\alpha}}{\alpha} \sqrt{\sin (\theta - \theta_{q})^{2}+q^{-2}\cos (\theta - \theta_{q})^{2}} \hspace{0.1cm}r^{\alpha}$ \\
\hline
\end{tabular}
\end{center}
\end{table}

\subsection{Separable potentials}

For the first class, we assign the lensing potential in a simple separable form:

\begin{equation}
\psi(r,\theta) = r^{\alpha} \ F(\theta, q, \theta_{q}) \ ,\label{psisep}
\end{equation}
where we have explicitly indicated the dependence  of the angular part
$F(\theta, q, \theta_{q})$  on the axial ratio $q$ ($0 < q < 1$) and
the position angle $\theta_{q}$ that specifies the orientation of the
lensing galaxy. The potentials (\ref{psisep}) are a generalization of
the pseudoisothermal elliptical potentials (PIEP) studied by
\cite{Kovner87b}.  This kind of model is in many aspects similar to
the pseudoisothermal elliptic mass distribution (PIEMD) models, such
as the singular isothermal ellipsoid (SIE) \cite{BlKoch}. However,
while PIEMD models can represent mass models with any elongation, PIEP
models represent physically plausible lenses only for some values of
the axial ratio $q$, such as $q>q_{0}$, being $q_0$ a suitable value
of the axial ratio of the isopotential contours
\cite{Kovner87b,Ka-Ko95}. For $q < q_{0}$, the elliptical isodensity
profile may in fact be boxy or disky. In particular, when dealing with
these elliptical models, we adopt the following expression for the
angular part:

\begin{equation}
F \propto \left [ \sin^{2} (\theta - \theta_{q})+q^{-2}\cos^{2}
(\theta
-
\theta_{q}) \right ]^{ 1/2},
\end{equation}
where $\theta_{q}$ is oriented from West to North. The angular part
$F$ also depends on a strength parameter b: for a singular isothermal
sphere ($\alpha=1$ and $F=$const) this parameter depends on the cosmological parameters and the
redshifts by means of a distance ratio, and on the central velocity
dispersion. In our case we can assume a similar dependence only for
{\it Model 3} (see Table \ref{tabella-potenziali-separabili}).

In Table \ref{tabella-potenziali-separabili} we give the expression of
$\psi$ for the three different models we consider. {\it Model 1}
is spherically symmetric, and  hence the shape function is
$F=\frac{b^{2-\alpha}}{\alpha}$.  On the other hand, both {\it Model
2} and {\it 3} are ellipticals, but for {\it Model 2} we fix $\alpha =
1$ as for isothermal mass distributions, while $\alpha$ is unfixed for
{\it Model 3}.

It is worth to note that, since the lensing potential $\psi$ and the
adimensional surface density $\kappa$ are related by a double
integration, the ellipticity of the isopotential contours is different
by that of isodensity ones. Therefore, we have to obtain a relation
between the axial ratio $q$ of the potential and the isodensity axial
ratio, named $q_{\kappa}$. For {\it Model 2}, the convergence is:

\begin{equation}
\kappa(r, \theta) =\frac{bq}{2 r} (\cos^{2} (\theta -\theta q)+q^{2} \sin^{2} (\theta -\theta q) )^{-\frac{3}{2}}
\ . \label{eq:conv-modello2}
\end{equation}
It is easy to see that $\kappa$ is always positive, i.e., {\it Model
1} is always correctly defined, and the analytical relation
$q_{\kappa}=q^{3}$ holds. Things get more complicated for {\it Model
3}. The convergence now turns out to be:

\begin{displaymath}
 \kappa(r, \theta)= \frac{b^{2-\alpha} r^{-2+\alpha}}{4q\alpha} [2(-1+q^{2}+\alpha^{2})\cos^{4}(\theta -\theta q)
\end{displaymath}
\begin{displaymath}
+q^{2} (2 (1+q^{2}(-1+\alpha^{2}))\sin^{4}(\theta -\theta q)+\alpha^{2}\sin^{4} 2 (\theta -\theta q)   )   ]
\end{displaymath}
\begin{equation}
\ \ \
 {\times} [( \cos^{2}(\theta -\theta q)+q^{2} \sin^{2}(\theta -\theta q)   )  ]^{-\frac{3}{2}}                \ .
\end{equation}
It is possible to see that, if one of the two factors
$q^{2}(\alpha^{2} - 1) + 1$ and $q^{2} + \alpha^{2} -1$ is
negative, then is $\kappa <0$. In particular, if $\alpha  >1$, we always have $\kappa >0$,
since the two quantities above are positive; instead, when $\alpha
<1$, the convergence $\kappa$ is negative if $q <
\sqrt{1-\alpha^{2}}$. For instance, if $\alpha = 0.5$ and $q < 0.866$,
the convergence is negative. The axial ratio $q_{\kappa}$ satisfies
the relation
\begin{equation}
q_{\kappa}=\bigg
(\frac{q(-1+q^{2}+\alpha^{2})}{-q^{2}+1+q^{2}\alpha^{2}}
\bigg)^{\frac{1}{2-\alpha}},
\end{equation}
that corresponds to more rapid trends for lower values of $\alpha$.

\subsection{Constant $M/L$ models}

The second class we consider contains models  that describe luminosity
profiles of elliptical galaxies. We use the de Vaucouleurs \cite{De
Vauc} and Hubble (REF) models (respectively {\it Model 4} and {\it 5}),
whose convergence $\kappa$ is given in Table  \ref{tabella-dens-M/L}.
We assume that the mass density is spherically symmetric, so that we
can avoid any difficulties in solving the lens equations for these
models. Given the surface density $\kappa$, the deflection angle is
easily obtained as \cite{SEF,Keeton}:

\begin{equation}
\hat{\alpha}(r) = \frac{2}{r} \int_{0}^{r}r' dr' \kappa(r'),\label{eq:angolo-lensing-simmetrico}
\end{equation}
and then the deflection potential is evaluated solving the equation $\hat{\alpha} = \nabla \psi$.

\begin{table}
\begin{center}
\caption{Mass distribution for constant $M/L$ models.
For the de Vaucouleurs model see Keeton \& Kochanek 1997.}
\label{tabella-dens-M/L}
\vspace{0.5cm}
\begin{tabular}{|c|c|}
\hline
\rule{0pt}{3ex}Model  & $\kappa$ \\
\hline
\hline
\rule{0pt}{3ex}Model 4  & $\frac{b}{R_{e}N'}e^{[-7.67(\frac{r}{R_{e}})^{\frac{1}{4}}]}$\\
\hline
\rule{0pt}{3ex}Model 5   & $\frac{1}{2} \frac{b^{2}}{s^{2}+r^{2}}$\\
\hline
\end{tabular}
\end{center}
\end{table}

\section{Single system modelling}\label{Single system}

Developing the general method used in CCRP01 and CCRP02 to model
quadruply lensed systems, we numerically solve systems of
nonlinear equations, imposing on the sample of  solutions some
criteria to select only the physical ones, that have to be
collected together  by means of an appropriate statistical
analysis.

\subsection{Method}
Here, we consider separable models with an assigned angular part and
more complex mass models for which it is necessary to assign the
surface density $\kappa$. In order to take into account all of the information
that come from a lensing event, we make use of time delay ratios, that
allow  to constrain the lensing parameters quite efficiently. We do
not use flux ratios since these may be contaminated by microlensing
\cite{ChRef79,Koop-deB2000}, and other effects, such as those due to
substructures in the lens galaxy \cite{Mao-Sch98}. Instead, time
delays are well measured for some quasars with great accuracy and are
less affected by spurious (and uncontrollable) effects. The unknown
parameters to be determined are the source position
$(r_{s},\theta_{s})$, the shear quantities ($\gamma,
\theta_{\gamma})$, the main lens model parameters (different for each
model), and the Hubble constant $H_{0}$. Actually, the higher
complexity of the models and the introduction of the time delays as
constraints lead to more complicated equations than those considered
in CCRP01 and CCRP02, and it is not possible to manipulate them to
reduce the set of lensing equations (\ref{timedelay}), (\ref{eqlens1})
and (\ref{eqlens2}) to a simpler form so as to speed up the
calculations. Let us write equations (\ref{eqlens1}) and
(\ref{eqlens2}) using the four images $i$, $j$, $k$ and $l$, and the
two equations coming from the time delay ratios among them, that are independent on h:

\begin{equation}
\frac{\Delta t_{ik}}{\Delta t_{ij}}=\frac{\Delta t_{ik}^{obs}}{\Delta t_{ij}^{obs} },   \qquad \frac{\Delta t_{il}}{\Delta t_{ij}}=\frac{\Delta t_{il}^{obs}}{\Delta t_{ij}^{obs} },
\end{equation}
where $\Delta t_{ij}^{obs}$, $\Delta t_{ik}^{obs}$ and $\Delta
t_{il}^{obs}$ are  the measured time delays. The
introduction of these two other equations permits to rise the
equations number, allowing to give a different constraint by the
images positions that does not depend on the first derivative of
the potential, but only on the potential.

In summary, we consider the observables without the errors,  using
their mean values: while for the image positions this assumption seems
immediately justified, for the time delays the uncertainties can be
considerably large. In a next section, we will show more accurately
which this assumption is reasonable.

We have a number of 10 equations  (8 from lens equations and the 2 due
to the time delay ratios), that we combine in a useful manner to
reduce their number to the unknowns number. For example, in the case
of $Model\; 1$ the unknowns are 6, i.e., $r_{s}$, $\theta_s$,
$\gamma$, $\theta_{\gamma}$, $b$, and $\alpha$, with in addition the
Hubble constant. We have to subtract the 8 lens equations to reduce
their number to 4, and we then add the two time delay ratios to
complete the system of 6 equations in 6 unknowns. We use the same
procedure for each model to obtain a system of $n$ equations and $n$
unknowns. We can numerically solve this system of  $n$ equations to
obtain $n$ unknown parameters (for our models $n$ is less than 10).
Each result of the solution of the system will be a set of $n$ values.
When we go to solve the system, we do not obtain a single solution,
but a set of solutions, since the system is highly non linear; these
solutions have to be selected by means of some selection criteria in
order to obtain values of parameters which give rise to physically
plausible models.

The search for the roots of the system begins with a choice of a range
of parameter values: one must be sure that there are no roots outside
the chosen range, and also a very large interval does not necessarily
include all roots, increasing CPU time. We choose a physically
acceptable interval for parameters, giving $\mathcal{N}$ random
starting points to the algorithm, where $\mathcal{N}$ is a number
fixed by the user\footnote{$\mathcal{N}$ should be large enough  to
explore a wide region in the parameter space, but not too large so as
to minimize CPU time. A possible strategy is to fix $\mathcal{N}$ to a
suitable value (for example, 2000 or 10000) and run the code more than
one time.}. From these starting points the CPU tries to converge to
the solutions, using the Newton's method. In particular, to assign the
same probability to the starting values of a parameter in the relative
range, we generate these values uniformly distributed, also to avoid
introducing any bias in the search.

We have developed a simple code, named LePRe\footnote{`Lepre' is the
Italian word for `hare'.}, Lensing Parameters Reconstruction, written
for {\it Mathematica}. Generally, solving the equation system yields
$\mathcal{M} <
\mathcal{N}$ solutions, because for some values of the starting points
Newton's method fails to converge to a  solution.

\subsection{Selection criteria}
These $\mathcal{M}$ solutions are not all physically acceptable and to
sort among these we have to impose a set of selection criteria, that
constrains parameter values to physically plausible ranges.
Schematically, we can describe these selection criteria as following.

\begin{enumerate}

\item{$0<r_{s}< max\{r_{i},r_{j},r_{k},r_{l}\}$: a lens will not produce images
arbitrarily  far away from the center of the lens; for large
values of $r_{s}$, there will be one image only at $(r,\theta) =
(r_{s},\theta_{s})$, and possibly others near the center of the
lens; in particular, we impose this cut because if the source is
outside the ring delineated by the most distant of the images it
is not possible to generate a 4-images configuration\footnote{One
can verify it using the web tool developed by K. Ratnatunga, which
generates the images of a lens system once given the lensing
potential, the source coordinates and the observational
characteristics (see
$\texttt{http://mds.phys.cmu.edu/ego\_cgi.html}$).}.}

\item{$0<\gamma <\gamma_{crit}$: the shear magnitude is positive by definition.
For the separable models, we choose as upper limit $\gamma \approx
\gamma_{crit}$, where $\gamma_{crit}$ is defined such that for
values of $\gamma \geq \gamma_{crit}$ the estimated $H_{0}$ becomes null,
and depends on the particular lens system to be considered
\cite{Wucknitz2002}. For PG 1115+080 is $\gamma_{crit}=0.22$, instead for
RX J0911+0551 is $\gamma_{crit}=0.56$. For axially symmetric lenses we
cannot fix an upper limit (since the hypothesis made in \cite{Wucknitz2002}
does not work), but in the analysis it is possible to obtain estimated
value of $\gamma$ higher than the one obtained for an elliptical separable
model.}

\item{\emph{The shear is approximately well directed}: we mean that the position
angle $\theta_{\gamma}$ must be directed towards the external mass
disturbance; e.g., if the shear were due to an external group of
galaxies, then $\theta_{\gamma}$ should be aligned with the cluster
mass centre, since there are no reasons why it should point elsewhere.
For the axially symmetric models this cut is not necessary because an
exact value of $\theta_{\gamma}$ is selected automatically from the
system solution to account for image configuration. Instead, for an
elliptical model, we have to take into account the degeneracy existing
between external asymmetry (the shear $\gamma$ and relative
orientation $\theta_{\gamma}$) and internal asymmetry provided by
axial ratio $q$ and angle $\theta_{q}$. Quantitatively, one could
accept only values comprised in the range $(\theta_{G\,
min},\theta_{G\, max})$, where $\theta_{G\, min}$ and
$\theta_{G\,max}$ are respectively two extreme galaxies that bound the
external group of galaxies.}

\item{\emph{The elliptical  profile of the galaxy has to be physically plausible}:
we assume $q_{0} < q < 1$, where $q_{0}$ is a particular value of $q$
such that, for fixed value of $\alpha$, the isodensity profile is
elliptical or nearly elliptical, and not strongly boxy; so, we avoid
those kinds of solutions to select the physically plausible one.}

\item{\emph{The range for other galaxy model parameters is chosen to have plausible
values of $\kappa$}. For {\it Model 1} and {\it Model 3} we must have
$0 < \alpha < 2$; as a matter of fact, the surface mass density scales
as $r^{\alpha -2}$, so that $\alpha < 2$ is needed in order to have
$\kappa$ monotonically decreasing with $r$. On the other hand, we
consider $\alpha> 0$ in order for the projected mass inside $r$ (that
scales as $r^{\alpha}$) to be reasonable. Since  $\psi >0$, we have to
assume the condition $b>0$ for the strength parameter. }

\item{\emph{Plausible index ind($\kappa$)}. The index is the logarithmic
derivative of $\kappa$, i.e., $ind(\kappa) \equiv \frac{d \log
\kappa}{d \log r}$. A lower bound can be fixed considering that
$\kappa$ has to be higher than the luminous profile; an upper
bound for PG 1115+080  is fixed following \cite{Wi-Saha2000},
i.e., $ind(\kappa)<-0.5$. For power law model this selection
criterium reduces to $\alpha<1.5$. }

\item{\emph{The set of parameters so found solves lens equations and time delay
ratios equations}. We insert the solutions in solved equations, to
check if the solution found is the correct one or a result from wrong
convergence of the numerical algorithm. We impose that the values of
parameters so found verify the equations within an accuracy fixed by
the user.}

\item{$h_{min}<h<h_{max}$. From the three time delays we obtain three values of
$h$ ($h_{ij}$, $h_{ik}$ and $h_{il}$)\footnote{We estimate three different $h$ to give more freedom to the obtained solution and to take into account that we are using  a numerical method.}, and  eliminate all the
solutions such that the predicted values of these three $h$'s differ
more than an $\epsilon$ ($=0.1$) from each other. Then, we estimate $h
\equiv (h_{ij}+h_{ik}+h_{il})/3$, selecting only those solutions which
give rise to values of $h$ in the range $(h_{min}, h_{max})$, being
these two parameters fixed by the user.  An upper bound is given by
age of globular clusters and by the limits from nucleochronology,
which indicate an age of the Universe of $t_{0}>7.8 \ Gyr$; instead, a
lower bound on $h$ can be given by $t_{0}\approx 20 \ Gyr$, since we
do not observe stellar systems with ages greatly exceeding this value.
In particular, to be conservative, we choose
$(h_{min},h_{max})=(0,2)$.}

\end{enumerate}
Obviously, one can also change the order of the constraints, and
remove or add some of these. At the end of  the application of such
selection criteria, and after having run the code more times, we get a
set of $\mathcal{L}$ solutions. It is  clear that there is a family of
different combinations of the parameters, that is consistent with the
observables and the selection criteria.

\subsection{Statistical interpretation}

It is possible to interpret the final sample of the solutions for
each parameter using the concept of Bayesian probability. In this
particular framework, the lensing parameters and $H_{0}$ represent
the ``variables'' of the ``system'', that are linked by a series
of ``relationships'', that need not be single-valued, i.e., the
lens mapping and time delay equations. We can associate to the
``variables'' a single value function, that is usually named
``information'', or more familiarly ``probability density
function'', which indicates how ``likely'' the particular
combination of the parameters is. Our final sample of solutions,
collected into the histograms, determines all the amount of
``information'' that we obtain about the ``system'' itself. This
sample is a subset of the parameter space, that contains all we
have to know about the ``system''. To extract a final result, we
have to select a better estimate of each parameter, making a
reasonable choice. We could use a figure of merit as made by
\cite{SW97}, but here we use the mean value of all the collected
values as final estimate for each parameter, instead for the
uncertainty on this estimate we think to be conservative in the
choice of the 68\% confidence level, i.e., the range in which the
68\% of the solutions is included. In the first place, this choice
works well in the simulations. Then, we argue that our choice
allows to take into account the weight of each single result in
the sample. When the distribution of the values is symmetric,
other choices as the maximum or the median are equivalent, but in
some cases the histograms have a degree of asymmetry, different
for each lensing model and quasar system, due to the particular
configuration of the images, the degeneracies and the available
data.

As we said before, by means of a simulated system, we will see that
this estimate of the uncertainty in the parameter determinations in
fact allows us to recover the values of these ones.

\section{A `two-system' model}\label{two-lens-model}
In a recent paper, Saha \& Williams extract an estimate of $H_{0}$ by means
of the joined use of more lensed systems \cite{SW2004}. Here, we show how
it is possible to build a `two-system' model fitting two lens systems with
general lensing potentials and a shared $H_{0}$. In order to obtain a
simpler solution to this problem we use an elliptical potential with a not
fixed angular part $F_{unk}(\theta)$ and external shear
\begin{equation}
\psi=F_{unk}(\theta)r^{\alpha}+\psi_{shear},
\end{equation}
already used in CCRP02. As shown in that paper, if we do not
assign the angular part of the potential, it is possible to write
the time delay between two images i and j in a simple form and, in
particular, for $h_{ij}$ the relation
\begin{eqnarray}
\lefteqn{ h_{ij} =\frac{1}{\Delta t_{ij}^{obs}}\frac{\tau_{100}}{2
\alpha}[(\alpha -2)(r_{i}^{2}-r_{j}^{2})} \nonumber \\
& & +2(1-\alpha) r_{s} (r_{i}\cos (\theta_{i}-\theta_{s} )-r_{j}\cos
(\theta_{j}-\theta_{s} )) \nonumber \\ & & +(\alpha -2) \gamma
(r_{i}^{2}\cos 2(\theta_{i}-\theta_{\gamma} )-r_{j}^{2}\cos
2(\theta_{j}-\theta_{\gamma} ))],
\end{eqnarray}
holds, where the lens parameters related to the angular trend of the lens
galaxy do not appear, but only the source position, the shear, the
parameter $\alpha$, and the normalized Hubble constant $h_{ij}$. The choice
of this potential allows us not only to explore a wide range of models but
also to obtain equations with a more little number of unknowns and then
more simply solvable.

We impose that the $h's$ due to different pairs of images and the
two different systems (i.e.,  $h_{ij}^{(1)}$, $h_{ik}^{(1)}$,
$h_{il}^{(1)}$, $h_{ij}^{(2)}$, $h_{ik}^{(2)}$ and
$h_{il}^{(2)}$)\footnote{The indices at the exponents specifies
the lensed system.} are equal. We could solve a system of
equations of the form $h_{ij}^{(1)}=h_{ik}^{(1)}, \,
h_{ij}^{(1)}=h_{ij}^{(2)},\, h_{ij}^{(2)}=h_{ik}^{(2)},\,.....$,
in order to obtain the 10 parameters
$r_{s}^{(1)},\theta_{s}^{(1)},\gamma^{(1)} ,
\theta_{\gamma}^{(1)},
\alpha^{(1)},r_{s}^{(2)},\theta_{s}^{(2)},\gamma^{(2)} ,
\theta_{\gamma}^{(2)},\alpha^{(2)}$. We verified, anyway, that it
is better to introduce a figure of merit of the form
\begin{equation}
\Gamma =
(h_{ij}^{(1)}-h_{ik}^{(1)})^{2}+(h_{ij}^{(2)}-h_{ik}^{(2)})^{2}+
(h_{ij}^{(1)}-h_{ij}^{(2)})^{2}+...,
\end{equation}
and we minimize it. We calculate the derivatives with respect to the 10
lens parameters and solve this system of 10 equations. Following the
previous section, we impose some selection criteria to select the physical
solutions. In this case, in order to reject many unphysical solutions we
have to impose a criterion not used previously: i.e., we require that the
source has a more constrained position\footnote{We impose the constraint on
the angle $\theta_{s}$}. This criterion is not arbitrary, since the source
can be located only in a little region to be able to generate a particular
configuration of the images (see \cite{SW2003}). After having obtained a
sample of solutions we have to calculate the six $h's$ by different pairs
of images and the two systems, and perform the mean of these values;
finally, we impose the constraint on $h's$ and obtain the final sample of
solutions, that gives out the estimates of the parameters.

We want to stress the fact that the algorithm discussed in this
Sec. and hence the developed method, has as main issue the
estimate of $h$. It does not allow to obtain precise estimate of
the position of the source and of the orientation of the external
perturbation, since we impose a strong constraint on them, primary
in order to obtain the correct estimate of $h$. As we will see
later, it also gives us a good determination of the other
parameters (i.e., $\gamma$, $\alpha$, and $h$).

We will compare the results with those obtained by using a similar
function of merit to model the single lens systems with the same
elliptical potentials. In this case the function $\Gamma$ is
simply modified introducing only the terms with the index $(1)$.

\section{Application to simulated systems}\label{Simulated systems}

In this section we verify the  correct working of the codes,
simulating real systems with image positions and time delays exactly
known (i.e., without observational uncertainties). Later on, analyzing
the system PG 1115+080, we will show that this last assumption is
reasonable, also for the real systems.

\subsection{Simulations}
For each simulation we fix the system parameters and  obtain the image
positions and  the three time delays by solving the lens equations
(\ref{eqlens1}) and (\ref{eqlens2}), and evaluating the time delay
function. We use these quantities as input for the codes in order to
verify if they are able to recover the correct values of the
originally fixed parameters.

\begin{table*}
\begin{center}
\caption{Simulated and estimated parameters, the angles are
written in radiants.}\label{tabella-simulazioni} \vspace{0.5cm}
\begin{tabular}{|@{}c@{}|@{}c@{}|@{}c@{}|@{}c@{}|@{}c@{}|@{}c@{}|@{}c@{}|}
\hline
\rule{0pt}{2.4ex} $$ & {\it Model 1} & {\it Model 2} & {\it Model 3}   & {\it Model 4} & {\it Model 5} \\
\hline
\hline
\rule{0pt}{2.4ex} $r_{s}^{est}\hspace{0.08cm}$ & $0.09_{-0.02}^{+0.02}$  & $0.4_{-0.02}^{+0.01}$ & $0.20_{-0.07}^{+0.03}$ &  $0.21_{-0.02}^{+0.02}$ & $0.15_{-0.01}^{+0.02}$\\
\hline
\rule{0pt}{2.4ex} $r_{s}\hspace{0.08cm}$ & $0.09$  & $0.4$ & $0.20$ &  $0.20$ & $0.15$\\
\hline
\rule{0pt}{2.4ex} $\theta_{s}^{est}\hspace{0.08cm}$ & $6.98_{-0.01}^{+0.01}$ & $2.48_{-0.06}^{+0.02}$ & $2.701_{-0.004}^{+0.004}$ & $5.00_{-0.01}^{+0.01}$ & $4.5_{-0.002}^{+0.002}$\\
\hline
\rule{0pt}{2.4ex} $\theta_{s}\hspace{0.08cm}$ & $6.98$ & $2.5$ & $2.7$ & $5.00$ & $4.5$\\
\hline
\rule{0pt}{2.4ex} $\gamma^{est}$ & $0.2_{-0.04}^{+0.04}$ & $0.19_{-0.09}^{+0.02}$ & $0.15_{-0.08}^{+0.01}$ &  $0.16_{-0.02}^{+0.02}$ & $0.15_{-0.01}^{+0.02}$\\
\hline
\rule{0pt}{2.4ex} $\gamma$ & $0.2$ & $0.2$ & $0.15$ &  $0.15$ & $0.15$\\
\hline
\rule{0pt}{2.4ex} $\theta_{\gamma}^{est}\hspace{0.08cm}$ & $3.77_{-0.01}^{+0.01}$ & $3.01_{-0.02}^{+0.09}$ & $6.32_{-0.12}^{+0.03}$ & $1.67_{-0.01}^{+0.01}$ & $3.8_{-0.001}^{+0.004}$\\
\hline
\rule{0pt}{2.4ex} $\theta_{\gamma}\hspace{0.08cm}$ & $3.77$ & $3$ & $6.34$ & $1.66$ & $3.8$\\
\hline
\rule{0pt}{2.4ex} $q^{est}$ & $-$ & $0.89_{-0.04}^{+0.01}$ & $0.9_{-0.01}^{+0.01}$  & $-$ & $-$\\
\hline
\rule{0pt}{2.4ex} $q$ & $-$ & $0.9$ & $0.9$  & $-$ & $-$\\
\hline
\rule{0pt}{2.4ex}  $\theta_{q}^{est}\hspace{0.08cm}$  & $-$ & $2_{-0.4}^{+0.1}$ & $2.29_{-0.11}^{+0.05}$  & $-$ & $-$\\
\hline
\rule{0pt}{2.4ex}  $\theta_{q}\hspace{0.08cm}$  & $-$ & $2$ & $2.3$  & $-$ & $-$\\
\hline
\rule{0pt}{2.4ex}  $b^{est}$ & $0.81_{-0.11}^{+0.08}$ & $1.8_{-0.1}^{+0.1}$ & $1.6_{-0.4}^{+0.7}$  & $1.76_{-0.57}^{+0.84}$ & $0.95_{-0.21}^{+0.06}$\\
\hline
\rule{0pt}{2.4ex}  $b$ & $0.8$ & $1.8$ & $1.5$  & $1.5$ & $1$\\
\hline
\rule{0pt}{2.4ex} $\alpha^{est}$ & $0.99_{-0.16}^{+0.17}$ & $-$ &  $1_{-0.1}^{+0.4}$ &  $-$ & $-$\\
\hline
\rule{0pt}{2.4ex} $\alpha$ & $1$ & $-$ &  $1$ &  $-$ & $-$\\
\hline
\rule{0pt}{2.4ex} $s^{est}, \hspace{0.01cm}R_{e}^{est}\hspace{0.01cm}$ & $-$ & $-$ & $-$   & $0.51_{-0.27}^{+0.45}$ & $0.17_{-0.14}^{+0.05}$\\
\hline
\rule{0pt}{2.4ex} $s, \hspace{0.01cm}R_{e}\hspace{0.01cm}$ & $-$ & $-$ & $-$   & $0.6$ & $0.2$\\
\hline
\rule{0pt}{2.4ex} $h^{est}$ & $0.69_{-0.18}^{+0.05}$ & $0.69_{-0.04}^{+0.08}$ & $0.67_{-0.22}^{+0.20}$ &  $0.66_{-0.10}^{+0.10}$ & $0.73_{-0.03}^{+0.08}$\\
\hline
\rule{0pt}{2.4ex} $h$ & $0.7$ & $0.7$ & $0.7$ &  $0.7$ & $0.7$\\
\hline
\end{tabular}
\end{center}
\end{table*}

In these simulation we adopt a flat homogeneous universe with cosmological constant, fixing:
\begin{equation}
(\Omega _{m}, \Omega_{\Lambda}, \Omega_{k}, h) = (0.3, 0.7, 0.0, 0.7) \ , \label{eq:parcosm}
\end{equation}
and:
\begin{equation}
(z_{d},z_{s})=(0.31,1.722),
\end{equation}
giving $\tau_{100}=33.37 \hspace{0.1 cm} \textrm{days} \hspace{0.1 cm} \textrm{arcsec}^{-2}$.

In addition to the expected solution, that is recovered with high
accuracy, the application of the codes generates other solutions,
different by the first one, due to the existence of the degeneracies
among the lensing parameters, that we will describe in the next
Section. The statistics on this solutions generates our estimate of
each parameter. Therefore, the uncertainties of the parameters in the
simulations and the real cases are not generated by a sort of
propagation of the errors of the observables, but are instead the
result of the degeneracies.

The application of the method to simulated systems allows to verify how
well it works in recovering the lens parameters and the normalized Hubble
constant h. We resume the results in Table \ref{tabella-simulazioni}, where
we report the simulated values of the parameters and the estimated ones.

\begin{figure}
\centering
\resizebox{8.5cm}{!}{\includegraphics{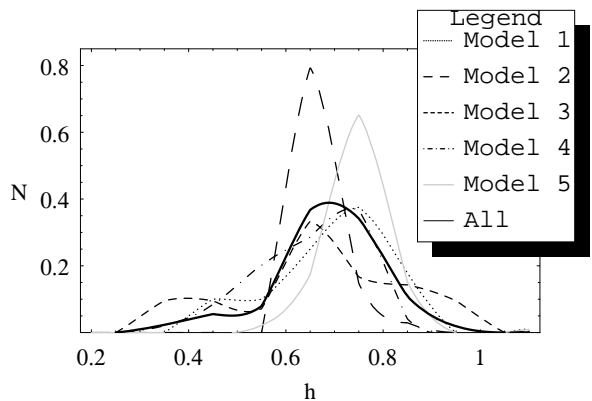}}
\caption{Distributions of the recovered values of the Hubble
constant. N indicates a number of values normalized to the total
number of results obtained for each model. This definition for $N$
will be used for all the following continuous distributions. The
distributions are obtained interpolating the recovered
histograms.} \label{ist-h-sim-totale}
\end{figure}

Finally, we show the distributions of the values of $h$ for the 5 models,
adding the total distribution, obtained averaging the single distributions,
in Figure \ref{ist-h-sim-totale}. In particular, we can  obtain a
``marginalized'' estimate of $h$ averaging the distributions, taking into
account all the recovered values in the distributions, and giving them
finite weights: if $N_{i}(h)$ indicates the distribution of the i-th model,
the combined $N(h)$ is given through a mean; this final distribution takes
into account the different weights for the different models.

We obtain the global result $h_{est}=0.69 {\pm} 0.11$. We may in fact consider
this procedure a way to obtain a final estimate of the Hubble constant
unaffected by spurious uncertainties of the single codes, having taken into
account different kinds of models and the whole parameters space. In this
way, we can see that the final estimate of $h$ is affected by a lower
uncertainty.

We have to verify whether the algorithm for the model with a common $h$ is
able or not to recover the lens parameters and, in particular, the Hubble
constant. The results are reported in Table
\ref{table-2_systems-simulation}: it is possible that some parameters are
not perfectly recovered, but the main ones are obtained with good accuracy.
\begin{table}
\begin{center}
\caption{Simulated and estimated parameters for the model with a not fixed angular part ({\it
Two-systems model}). }\label{table-2_systems-simulation}
\vspace{0.5cm}
\begin{tabular}{|c|c|c|}
\hline
\rule{0pt}{2.4ex} $$ & $\textrm{simulated}$ & $\textrm{estimated}$ \\
\hline \hline
\rule{0pt}{2.4ex} $r_{s}^{(1)}$  & $0.25$ & $0.28_{-0.14}^{+0.13}$ \\
\hline
\rule{0pt}{2.4ex} $\theta_{s}^{(1)}$  & $1.6$ & $1.66 _{-0.49}^{+0.33}$ \\
\hline
\rule{0pt}{2.4ex} $\gamma^{(1)}$  & $0.13$ & $0.13 _{-0.06}^{+0.04}$ \\
\hline
\rule{0pt}{2.4ex} $\theta_{\gamma}^{(1)}$  & $3.00$ & $2.98 _{-0.14}^{+0.31}$ \\
\hline
\rule{0pt}{2.4ex} $\alpha^{(1)}$  & $1.1$ & $1.13_{-0.13}^{+0.14}$ \\
\hline
\rule{0pt}{2.4ex} $r_{s}^{(2)}$  & $0.15$ & $0.24_{-0.16}^{+0.18}$ \\
\hline
\rule{0pt}{2.4ex} $\theta_{s}^{(2)}$  & $2.8$ & $2.90_{-0.36}^{+0.44}$ \\
\hline
\rule{0pt}{2.4ex} $\gamma^{(2)}$  & $0.15$ & $0.13 _{-0.08}^{+0.03}$ \\
\hline
\rule{0pt}{2.4ex} $\theta_{\gamma}^{(2)}$  & $3.00$ & $2.96_{-0.30}^{+0.06}$ \\
\hline
\rule{0pt}{2.4ex} $\alpha^{(2)}$  & $1$ & $1.06_{-0.09}^{+0.19}$ \\
\hline
\rule{0pt}{2.4ex} $h$  & $0.7$ & $0.68_{-0.14}^{+0.13}$ \\
\hline
\end{tabular}
\end{center}
\end{table}
In addition, we verify the correct working of the code when we fit the same
potential to a single simulated system. The results are reported in Table
\ref{table-1_system-simulation}. We note that the estimate of $h$ for the two-system model has a lower uncertainty than that of the single modelling

\begin{table}
\begin{center}
\caption{Simulated and estimated parameters for the model with $F(\theta)$ unknown, fitted to a single system. }\label{table-1_system-simulation}
\vspace{0.5cm}
\begin{tabular}{|c|c|c|}
\hline
\rule{0pt}{2.4ex} $$ & $\textrm{simulated}$ & $\textrm{estimated}$ \\
\hline \hline
\rule{0pt}{2.4ex} $r_{s}^{(1)}$  & $0.25$ & $0.25_{-0.18}^{+0.19}$ \\
\hline
\rule{0pt}{2.4ex} $\theta_{s}^{(1)}$  & $1.6$ & $1.70 _{-0.47}^{+0.28}$ \\
\hline
\rule{0pt}{2.4ex} $\gamma^{(1)}$  & $0.13$ & $0.12 _{-0.10}^{+0.06}$ \\
\hline
\rule{0pt}{2.4ex} $\theta_{\gamma}^{(1)}$  & $3.00$ & $2.96 _{-0.33}^{+0.53}$ \\
\hline
\rule{0pt}{2.4ex} $\alpha^{(1)}$  & $1.1$ & $1.20 _{-0.22}^{+0.27}$ \\
\hline
\rule{0pt}{2.4ex} $h$  & $0.7$ & $0.66_{-0.18}^{+0.19}$ \\
\hline
\end{tabular}
\end{center}
\end{table}

\subsection{Model reconstruction from other lens models}
It is well known that the main uncertainty in applying the gravitational
lensing in order to estimate the Hubble constant and to investigate the
astrophysical properties of the galaxies comes from the lack of knowledge
of the ``true'' galaxy model (see, for instance, \cite{SEF}). Here, we want
briefly to analyze the biases and the errors due to the use of an
``incorrect'' parametric model to fit the observable data. A way to face
this important question consists in to build simulated systems for a lens
model and then to fit the image positions and time delays generated by it,
using an other functional form, studying the change in the lens parameters
and above all in $h$. We can furnish a qualitative estimate of the effect
of the model dependence on $H_0$. At the same time the procedure allows to
quantify this ``systematic'' errors.

By means of a simulated system built using the Hubble model, we
fit to the image positions and the time delays so obtained the
{\it Model 1} and {\it Model 2}. The analysis of the most
significant parameters shows interesting trends which are
partially already well known. In particular, in our simulated
system $h=0.7$, and the fitting of the other models gives us the
means values $h=0.35$ for the {\it Model 1} and $h=0.46$ for the
{\it Model 2}, respectively with a percentage change of $50\%$ and
$34\%$. This percentage obviously changes if we simulate other
lens systems.  This shows that if the ``correct'' model for a lens
is one with constant mass-to-light ratio and it tries to shape it
with a separable model we obtain a lower estimate of $h$ than that
obtained with the first one; this verifies some previous results
in literature (see \cite{Koch02} for an analysis made on real
systems). Similar trends are obtained if we create a simulated
system using a de Vaucouleurs model. It is also possible to
analyze the uncertainties introduced by the lack of the internal
ellipticity of the lens galaxy. If we try to fit with the {\it
Model 1} the observables generated with the {\it Model 2} we
obtain a lower mean $h$ but if we consider the errors this
estimate is in agreement with the simulated value, instead the
estimated value for $\alpha$ raises.

\section{Parameter degeneracies}\label{correlations}

By means of simulated systems we can also obtain statistical
correlations among parameters in order to investigate the
degeneracies among them and to study the effect of varying each
parameter on the final Hubble constant estimate. Below, we discuss the
results we found for each model.

\begin{itemize}

\item{\textbf{Model 1.} We see that $r_{s},\gamma, b$, and $h$ correlate each other,
and anticorrelate with $\alpha$. In particular, we verify the scaling
laws \cite{Wucknitz2002}:

\begin{equation}
r_{s} \propto 2-\alpha, \hspace{0.25 cm} \gamma \propto 2-\alpha
\hspace{0.25 cm}, \hspace{0.25 cm} h \propto 2-\alpha .
\end{equation}

Wucknitz \& Refsdal 2001 give a simple interpretation for these
scaling laws in terms of mass-sheet degeneracy. In Figure
\ref{Fit-h-alfa-potalfa} we report as an example the correlation
$h-\alpha$, fitting the line $h=a(2-\alpha)$, where $a$ is a
proportionality constant.}

\begin{figure}
\centering
\resizebox{8.5cm}{!}{\includegraphics{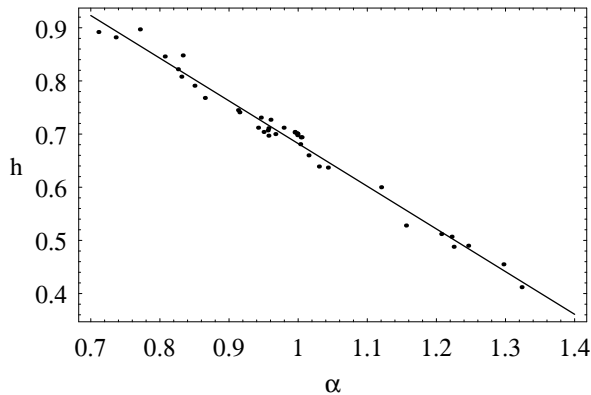}}
\caption{Scaling law $h \propto 2-\alpha$ for {\it Model 1}.} \label{Fit-h-alfa-potalfa}
\end{figure}

\item{\textbf{Model 2.} For this isothermal model we observe that $r_{s}, b$, and $h$
correlate among each other, and anticorrelate with $\gamma$ and $q$;
almost all of these are linear correlations with a high correlation
degree. We note (see also later considerations about correlations for
{\it Model 3}) that the absence of the parameter $\alpha$ changes the
correlations among the shear $\gamma$ and the other parameters, i.e.,
since $\alpha$ is fixed, $\gamma$ now correlates negatively with
$r_{s}, b$, and $h$, and not positively.}

\item{\textbf{Model 3.} $r_{s}, \gamma, q, b$, and $h$  correlate among each other and
anticorrelate with $\alpha$. Here, the positive correlation among
$\gamma$ and the other parameters comes back: this allows us to
confirm that, in {\it Model 2}, it is the condition $\alpha=1$ that
changes the correlations of the external shear and not the presence of
an intrinsic ellipticity. We also note the obvious positive
correlation between $q$ and $\gamma$ (as for {\it Model 2}). Higher
values of the internal ellipticity (i.e., lower values of $q$) require
 lower values of $\gamma$.}

\item{\textbf{Model 4.} For the de Vaucouleurs model we find similar linear
correlations: $r_{s}, \gamma, b$, and $h$ correlate among each other
and anticorrelate with $R_{e}$, which now assumes the role of
$\alpha$. For example, for higher values of $R_{e}$, this model
predicts lower values of Hubble constant.}

\item{\textbf{Model 5.} For the Hubble model we find the same correlations of the preceding model, simply replacing $R_{e}$ with the core radius $s$}

\end{itemize}

For real systems it is more difficult to obtain these correlations.
Therefore, we do not discuss the results for them, and we prefer to
resume the obtained  dependencies in the following:

\begin{itemize}
\item{The parameters $r_{s}$, $b$, and $h$ always correlate positively among each
other; i.e., more massive lenses require higher values of the radial
position of the source and of the Hubble constant. The observations
show that more luminous galaxies give larger angular separations of
the images, in agreement with our correlations: a larger $b$ generates
a larger $r_s$ and, hence, a larger separation of the images.}

\item{The parameters that determine the radial profile of the lens galaxy ($\alpha$, $R_e$, or $s$) correlates negatively with the other ones.}

\item{Except for the {\it Model 2}, all the models reveal a positive correlation of
the shear with $r_{s}$, $b$, and $h$.}

\end{itemize}

\section{Application to PG 1115+080 and RX J0911+0551} \label{applications}

Having checked that the numerical codes indeed work correctly
recovering the values of the lensing parameters and Hubble
constant, we apply them to real quadruply imaged systems. We need
a four images system with a good astrometry of the lensing galaxy
and image positions, and time delays measured with high accuracy.
Then, we also need that there is a single galaxy acting as lens:
for instance, with our models we cannot study a gravitational lens
like B 1608+656, since there are two lensing galaxies; in this
case, a more complex (two) lens model should be used. There are
only three systems that satisfy all these requirements: PG
1115+080, RX J0911+0551, and B 1422+231. Here, we choose to apply
the codes only to PG 1115+080 and RX J0911+0551, since B 1422+231
has time delays measured with a high uncertainty (see, for
instance \cite{PN01}). We will also combine the results from the
two systems to get a better estimate of the Hubble constant. We
adopt a flat cosmology with $(\Omega_{m}, \Omega_{\Lambda}) =
(0.3, 0.7)$, discussing for each system the influence of changing
the cosmological parameters on the final estimate of $H_{0}$.

\subsection{Application to PG 1115+080}

PG 1115+080 was discovered by \cite{Weymann80}, originally  as a
triple lensed quasar. Later, it has been possible to split an image
(the image A) in two components, $A_1$ and $A_2$. Therefore, this
system consists of four images ($A_1$, $A_2$, $B$, and $C$) of a radio
quiet QSO at $z_{s}=1.722$, with an elliptical galaxy as lens
belonging to a group of galaxies at $z_{d} = 0.31$. This group,
situated at South-West, contains $\sim$ 11 galaxies, with a luminous
centroid at $(r_{g},\theta_{g}) = (20'' {\pm} 0.2'',-117^{o}{\pm} 3^{o})$.
Iwamuro et al. fitted the luminous profile of the lensing galaxy
with a de Vaucouleurs model with $R_{e} = 0''.58 {\pm} 0''.05$, and
measured an ellipticity $\sim 0.1$ and a position angle of $\sim
65^{0}$ from north \cite{Iwamuro2000}.

Here, we use  image coordinates measured by \cite{Impey-et-al98}, and
the time delay obtained by \cite{Barkana97}. In particular, in
\cite{Barkana97} it measured a time delay between the components $B$ and
$C$ of $\Delta t_{BC} = 25.0 {\pm} 1.7$\,days, consistent with the
previous value $23.7 {\pm} 3.4$\,days from \cite{Schechter97}. By
contrast, the time delay ratio $r_{ABC}=\Delta t_{AC}/ \Delta
t_{BA}=1.13^{+0.18}_{-0.17}$ found by Barkana is in conflict with the
value $0.7 {\pm} 0.3$ from \cite{Schechter97}.

We can fit this system with the models previously discussed and
observe some peculiar trends in the obtained parameter values,
that we discuss in the following:

\begin{itemize}

\item{The source positions are consistent with each other except for {\it Model 2}, that has a little disagreement in its mean value.}

\item{As expected, axially symmetric models predict a higher mean value of the shear, to account  for the lack of an internal asymmetry, being in agreement with the previous estimates of $\sim 0.1$. It is also interesting to note that $\theta_{\gamma}$ is perfectly oriented towards the luminous centroid of the external group.}

\item{The isopotential profile has a small ellipticity, which increases
when we go to consider the relative isodensity profiles. In
particular, {\it Model 2} predicts an E0\,-\,E1 galaxy, while {\it
Model 3} is also consistent with more elliptical profiles. Instead,
$\theta_{q}$  marginally agrees with the luminous profile orientation,
but is nonetheless in agreement with results obtained with other
techniques \cite{Impey-et-al98}.}

\item{We obtain $\alpha$ values consistent (also if marginally) with the
nearly isothermal model, with major preference for {\it Model 3} (see \cite{Koch02}).}

\item{{\it Model 4} predicts a value of $R_{e}$ in agreement with
the measured one of $0''.58{\pm} 0''.05$ by \cite{Iwamuro2000}, while for {\it
Model 5} the core radius $s$ is very small, being consistent with a null
core within the uncertainties.}

\item{At least, the most important result concerns the estimated Hubble constants.
Constant $M/L$ models predict higher values than separable models, as
yet found previously in literature \cite{Courbin97,Koch02}. Then, for
the isothermal {\it Model 2}, we verify a result of
\cite{Impey-et-al98} that gives a low value using a SIE, an isothermal
model similar to our model, but with a different angular part.}

\end{itemize}

\begin{figure*}
\centering \resizebox{18cm}{!}{\includegraphics[width=5cm,
height=0.9cm]{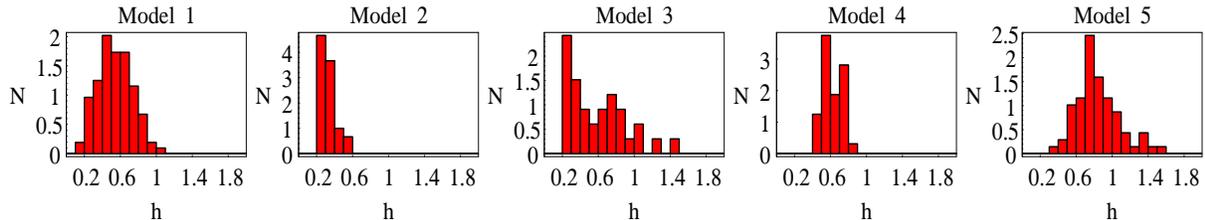}} \caption{Histograms of the
recovered values of the Hubble constant for PG 1115+080. The area
under the histograms is normalized to unity.} \label{ist-h-PG}
\end{figure*}

\begin{figure}
\centering
\resizebox{8.5cm}{!}{\includegraphics{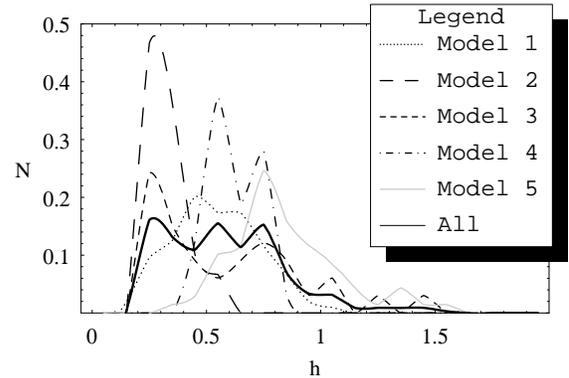}}
\caption{Distributions of all the recovered values of the Hubble
constant for PG 1115+080.} \label{distributions_PG}
\end{figure}

We report the results of the application of our procedure in Fig.
\ref{ist-h-PG}. In Fig. \ref{distributions_PG} we group together
the different distributions. The mean distribution gives the value
$H_{0}= 58 {\pm} 27 \ {\rm km \ s^{-1} \ Mpc^{-1}}$. The mean
distribution for the constant mass-to-light ratio models gives us
$H_{0}= 73 {\pm} 22 \ {\rm km \ s^{-1} \ Mpc^{-1}}$ in according
with the recent results in \cite{Koch02}. If we change the values
of cosmological parameters, our final estimate can change of $\sim
3 \%$, but the spread becomes $\sim 12\%$ if we also consider
inhomogeneous models.

\begin{table}
\begin{center}
\caption{Estimated parameters for PG 1115+080. $\theta_{q}$ is oriented from West to North.}
\vspace{0.5cm}
\begin{tabular}{|@{}c@{}|@{}c@{}|@{}c@{}|@{}c@{}|@{}c@{}|@{}c@{}|@{}c@{}|}
\hline
\rule{0pt}{2.4ex} $$ & {\it Model 1} & {\it Model 2} & {\it Model 3}   & {\it Model 4} & {\it Model 5} \\
\hline
\hline
\rule{0pt}{2.4ex} $r_{s}\hspace{0.08cm}('')$ & $0.14_{-0.05}^{+0.05}$  & $0.10_{-0.02}^{+0.02}$ & $0.16_{-0.09}^{+0.08}$ &  $0.17_{-0.02}^{+0.02}$ & $0.19_{-0.05}^{+0.08}$\\
\hline
\rule{0pt}{2.4ex} $\theta_{s}\hspace{0.08cm}(^{o})$ & $24_{-3}^{+3}$ & $18_{-6}^{+5}$ & $21_{-5}^{+5}$ & $24_{-4}^{+2}$ & $24_{-2}^{+2}$\\
\hline
\rule{0pt}{2.4ex} $\gamma$ & $0.15_{-0.05}^{+0.05}$ & $0.09_{-0.03}^{+0.02}$ & $0.13_{-0.10}^{+0.07}$ &  $0.16_{-0.03}^{+0.02}$ & $0.17_{-0.04}^{+0.04}$\\
\hline
\rule{0pt}{2.4ex} $\theta_{\gamma}\hspace{0.08cm}(^{o})$ & $-114_{-3}^{+2}$ & $-115_{-8}^{+15}$ & $-119_{-7}^{+11}$ & $-114_{-4}^{+2}$ & $-115_{-3}^{+3}$\\
\hline
\rule{0pt}{2.4ex} $q$ & $-$ & $0.97_{-0.03}^{+0.02}$ & $0.95_{-0.06}^{+0.04}$  & $-$ & $-$\\
\hline
\rule{0pt}{2.4ex} $q_{k}$ & $-$ & $0.91_{-0.08}^{+0.06}$ & $0.70_{-0.19}^{+0.25}$  & $-$ & $-$\\
\hline
\rule{0pt}{2.4ex}  $\theta_{q}\hspace{0.08cm}(^{o})$  & $-$ & $155_{-21}^{+41}$ & $180_{-22}^{+60}$  & $-$ & $-$\\
\hline
\rule{0pt}{2.4ex}  $b$ & $1.06_{-0.07}^{+0.05}$ & $0.98_{-0.10}^{+0.16}$ & $1.05_{-0.14}^{+0.19}$  & $1.51_{-0.41}^{+0.8}$ & $0.50_{-0.13}^{+0.20}$\\
\hline
\rule{0pt}{2.4ex} $\alpha$ & $0.62_{-0.44}^{+0.50}$ & $-$ &  $0.70_{-0.50}^{+0.70}$ &  $-$ & $-$\\
\hline
\rule{0pt}{2.4ex} $s, \hspace{0.01cm}R_{e}\hspace{0.01cm}('')$ & $-$ & $-$ & $-$   & $0.72_{-0.37}^{+0.48}$ & $0.11_{-0.10}^{+0.30}$\\
\hline
\rule{0pt}{2.4ex} $h$ & $0.54_{-0.21}^{+0.19}$ & $0.33_{-0.08}^{+0.09}$ & $0.59_{-0.30}^{+0.34}$ &  $0.62_{-0.08}^{+0.11}$ & $0.84_{-0.20}^{+0.25}$\\
\hline
\end{tabular}
\end{center}
\end{table}

\begin{figure}
\centering
\resizebox{7.5cm}{!}{\includegraphics{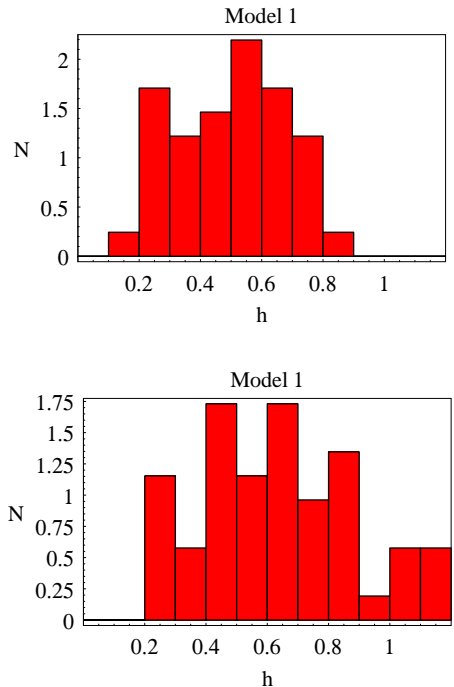}}
\caption{Histograms of the recovered values of the Hubble constant for PG 1115+080
with {\it Model 1} changing the time delays.}
\label{ist_time_delays_var}
\end{figure}

Finally, in order to verify the honesty of our assumption of
considering the observables without errors, we  proceed as
following: we consider as input values for the codes the mean
values of the observables, and then, we change the time delays of
an amount about the 15\%. For example, for the {\it Model 1} we
obtain the results in figure \ref{ist_time_delays_var}. The two
best values of h are: $h_{down}=0.50^{+0.16}_{-0.21}$ and
$h_{up}=0.64^{+0.23}_{-0.25}$, in agreement with each other and
with the value obtained using the central value. Therefore, the
uncertainty in the time delays is extensively absorbed by the
uncertainty due to the parameter degeneracies.

\subsection{Application to RX J0911+0551}

RX J0911+0551 was discovered by \cite{Bade97} as a quadruple
imaged QSO, in the ROSAT All-Sky Survey (RASS). The source QSO is
situated at $z_{s}=2.8$, while the lensing galaxy is at $z_{d} =
0.77$. The image configuration is peculiar: the three images
$A_1$, $A_2$ and $A_3$ are close to each other and to the lensing
galaxy ($A_1-A_2=0.478''$, $A_2-A_3=0.608''$), while image $B$ is
more distant than the lens, requiring a large external shear. The
only intrinsic ellipticity does not allow to account for this
configuration, and we need an external asymmetry. There is, in
fact, a nearby group at about $38''$ South-West, and at redshift
$z_{group}=0.7{\pm} 0.1$ that can be the origin of this external
perturbation. \cite{Hjorth02} measured the time delay obtaining
the results $\Delta t_{BA_{1}}=-143 {\pm} 6 \hspace{0.1 cm}
\textrm{days}$, $\Delta t_{BA_{2}}=-149 {\pm} 6 \hspace{0.1 cm}
\textrm{days}$ and $\Delta t_{BA_{3}}=-154 {\pm} 16 \hspace{0.1
cm} \textrm{days}$. It is also possible to observe a second galaxy
near the main one, that can affect the modelling.

In the following we present the main results of the  application
of our codes:

\begin{itemize}

\item{The models predict nearly the same values for the source positions, except {\it Model 3}, that provides a lower value.}

\item{The axially symmetric model yields a high value of the shear; for
{\it Model 2} we obtain $\gamma = 0.22$, while {\it Model 3} requires
a lower value. The shear angle points towards the external group,
allowing for the particular configuration observed.}

\item{The elliptical model requires a density distribution with high ellipticity
consistent with an E1\,-\,E5 galaxy, with a position angle oriented
towards the external group.}

\item{{\it Model 1} gives a lower value of $\alpha$ in agreement with the value
from \cite{Schechter2000}, while for the other elliptical model this
value is high to account for the low values of $\gamma$ and $q$.}

\item{The effective radius $R_{e}$ is low ($\sim 0.2$), and the core radius of the
Hubble model is consistent with zero.}

\item{The recovered value of $H_{0}$ are higher than those obtained using PG 1115+080, with dramatically
high uncertainties. The distributions of these values are too much
flat giving us a little quantity of statistical information; maybe,
using more complex models and other constraints we could improve these
results.}

\end{itemize}

Our method allows  to obtain the better estimate for the system
parameters, but this circumstance does not also allow us to claim that
a particular model fits the image positions and the other observables
with higher accuracy. We note that the axially symmetric models are
able to recover the correct position of the images $A_{1}$, $A_{2}$
and $A_{3}$, but do not allow to get the position of the fourth image
due to lacking of internal ellipticity.

\begin{table}
\begin{center}
\caption{Estimated parameters for RX J0911+0551.}
\vspace{0.5cm}
\begin{tabular}{|@{}c@{}|@{}c@{}|@{}c@{}|@{}c@{}|@{}c@{}|@{}c@{}|@{}c@{}|}
\hline
\rule{0pt}{2.4ex} $$ & {\it Model 1} & {\it Model 2} & {\it Model 3}  & {\it Model 4} & {\it Model 5} \\
\hline
\hline
\rule{0pt}{2.4ex} $r_{s}\hspace{0.08cm}('')$ & $0.60_{-0.14}^{+0.21}$  & $0.50_{-0.16}^{+0.16}$ & $0.29_{-0.17}^{+0.20}$ &  $0.52_{-0.12}^{+0.24}$ & $0.49_{-0.15}^{+0.19}$\\
\hline
\rule{0pt}{2.4ex} $\theta_{s}\hspace{0.08cm}(^{o})$ & $87_{-6}^{+6}$ & $88_{-6}^{+4}$ & $87_{-7}^{+8}$ & $85_{-7}^{+10}$ & $84_{-7}^{+5}$\\
\hline
\rule{0pt}{2.4ex} $\gamma$ & $0.42_{-0.09}^{+0.07}$ & $0.22_{-0.15}^{+0.07}$ & $0.17_{-0.13}^{+0.28}$ &  $0.42_{-0.08}^{+0.10}$ & $0.38_{-0.11}^{+0.12}$\\
\hline
\rule{0pt}{2.4ex} $\theta_{\gamma}\hspace{0.08cm}(^{o})$ & $171_{-3}^{+3}$ & $173_{-7}^{+11}$ & $181_{-22}^{+36}$ & $170_{-5}^{+4}$ & $168_{-6}^{+6}$\\
\hline
\rule{0pt}{2.4ex} $q$ & $-$ & $0.88_{-0.08}^{+0.08}$ & $0.80_{-0.15}^{+0.15}$  & $-$ & $-$\\
\hline
\rule{0pt}{2.4ex} $q_{k}$ & $-$ & $0.68_{-0.17}^{+0.20}$ & $0.68_{-0.20}^{+0.24}$  & $-$ & $-$\\
\hline
\rule{0pt}{2.4ex}  $\theta_{q}\hspace{0.08cm}(^{o})$  & $-$ & $88_{-42}^{+44}$ & $81_{-29}^{+48}$  & $-$ & $-$\\
\hline
\rule{0pt}{2.4ex}  $b$ & $0.96_{-0.08}^{+0.09}$ & $1.11_{-0.26}^{+0.23}$ & $0.94_{-0.56}^{+0.54}$  & $6.34_{-4.36}^{+5.33}$ & $0.38_{-0.12}^{+0.23}$\\
\hline
\rule{0pt}{2.4ex} $\alpha$ & $0.20_{-0.17}^{+0.38}$ & $-$ &  $1.42_{-0.90}^{+0.33}$ &  $-$ & $-$\\
\hline
\rule{0pt}{2.4ex} $s, \hspace{0.01cm}R_{e}\hspace{0.01cm}('')$ & $-$ & $-$ & $-$   & $0.18_{-0.14}^{+0.18}$ & $0.06_{-0.06}^{+0.36}$\\
\hline
\rule{0pt}{2.4ex} $h$ & $0.99_{-0.28}^{+0.40}$ & $1.00_{-0.40}^{+0.44}$ & $0.57_{-0.33}^{+0.52}$ &  $0.75_{-0.35}^{+0.41}$ & $0.83_{-0.46}^{+0.59}$\\
\hline
\end{tabular}
\end{center}
\end{table}

We may also try to apply an iterative procedure to verify the correct
working of the codes for this lens system. We fix the values of almost
all the parameters leaving two of them free. Then, we choose one of
those two, solving the lens equation relative to the image $B$ and
obtaining an estimate for the other parameter\footnote{We solve this
equation with 1 unknown, fixing as starting point the value obtained
using the code.}. After finding this value, we can iterate the process
solving the equation for the other parameter. After that, we again
calculate the image position without being able to recover the correct
ones. Therefore, we argue that \emph{is not possible to fit the image
positions with an axial symmetric model with external shear}.

Our results agree with \cite{Burud1998}; they show that the
particular image  configuration requires a minimum ellipticity for
the galaxy of $\epsilon_{min}=0.075$ and a minimum shear of
$\gamma_{min}=0.15$, applying the method of \cite{WM97}. In
particular, we predict very high values of $\gamma$, except for
{\it Model 2} and {\it Model 3}, that needs a lower mean value,
but in agreement with that lower bound.

\begin{figure*}
\centering \resizebox{18cm}{!}{\includegraphics[width=5cm,
height=0.9cm]{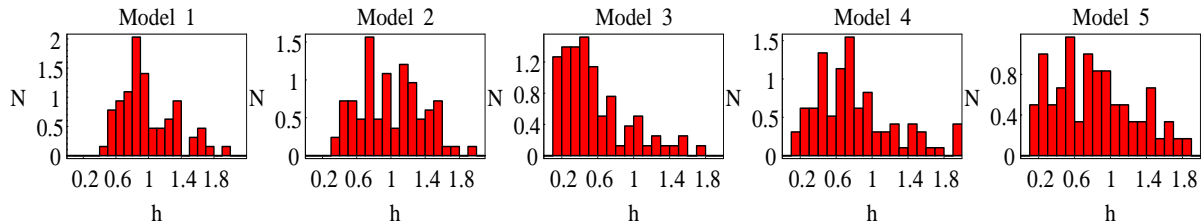}} \caption{Histograms of the
recovered values of the Hubble constant for RX J0911+0551.}
\label{ist-h-RX}
\end{figure*}

\begin{figure}
\centering
\resizebox{8.5cm}{!}{\includegraphics{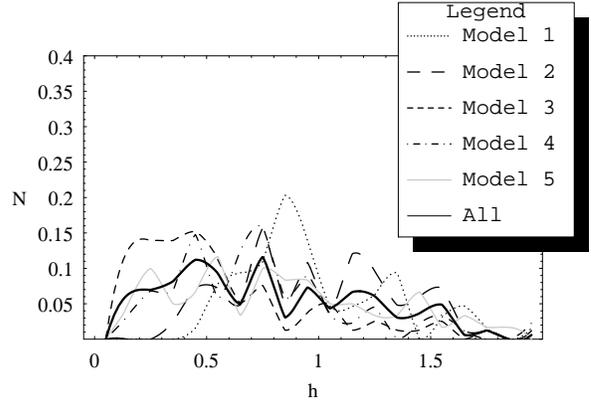}}
\caption{Distributions of all recovered values of the Hubble constant for RX J0911+0551.} \label{ist-h-RXtotale-ellittici}
\end{figure}

The histograms of the $H_0$ values are shown in Figs.
\ref{ist-h-RX}. We collect all the distributions in Fig.
\ref{ist-h-RXtotale-ellittici}. We give two final estimates of
Hubble constant, the first one only including the elliptical
models, and the second one including all models. Using the
elliptical models we obtain $H_{0}=81 {\pm} 41 \hspace{0.1 cm}
\textrm{km} \hspace{0.05 cm} \textrm{s}^{-1} \textrm{Mpc}^{-1}$,
instead taking into account all the models we have $H_{0}=77 {\pm}
43 \hspace{0.1 cm} \textrm{km} \hspace{0.05 cm} \textrm{s}^{-1}
\textrm{Mpc}^{-1}$. Using a power-law model with external shear
(the \mbox{`` Yardstick''} model) \cite{Schechter2000} gets a low
value of $\alpha$, in agreement with the value obtained for {\it
Model 1}, and $H_{0}=42 \hspace{0.1 cm} \textrm{km} \hspace{0.05
cm} \textrm{s}^{-1} \textrm{Mpc}^{-1}$, with an uncertainty
\mbox{$\sim 10-20 \%$}. This value does not agree with the result
obtained here because in \cite{Schechter2000} is used a time delay
of $200 \hspace{0.1 cm} \textrm{days}$ among the mean image $A$
and the image $B$, different by the one we use. Instead, including
in the model the main lens galaxy, the cluster of galaxy, and
individual galaxies in the cluster \cite{Hjorth02} get an estimate
of $H_{0}=71{\pm} 4 \hspace{0.1 cm} (\textrm{random}, 2 \sigma) {\pm}
8 \hspace{0.1 cm} (\textrm{systematic}) \hspace{0.1 cm}
\textrm{km} \hspace{0.05 cm} \textrm{s}^{-1} \textrm{Mpc}^{-1}$,
that agrees with the value we obtain.

Fixing other values  for cosmological parameters, the maximum spread
is $\sim 10 \%$, but considering a Dyer \& Roeder universe we have a
value $\sim 30-40 \%$, since the redshifts $z_{s}$ and $z_{d}$ of RX
J0911+0551 are higher than those of PG 1115+080. Actually, RX
J0911+0551 is not the ideal system to obtain an accurate estimate of
$H_{0}$, since there is a high uncertainty introduced by the arbitrary
choice of different cosmological parameters.

Finally, we remark that a simple elliptical potential with external
shear, also being able to account for image configuration, is an
approximate attempt to model a complex system as RX J0911+0551. Our
modelling, in fact, allows to obtain useful information about the
system, but we expect to obtain better results using, for example, a
SIS or SIE to describe the contribution of the external group, and not
an approximation as the external shear. Moreover, it is necessary to
take into account the contribution of the second galaxy and the other
objects in the group.

\subsection{Marginalized estimate of $H_{0}$}

For each systems we obtained an estimate of $H_0$ performing a mean of the
final distributions obtained fitting each model, and then deriving the mean
value and the uncertainty from it. Now, let us combine these
results\footnote{For RX J0911+0551 we only use the two elliptical models.},
multiplying the two final distributions. The final estimate turns out to
be\,:
\begin{equation}
H_{0}=56 {\pm} 23 \hspace{0.1 cm} \textrm{km} \hspace{0.05 cm}
\textrm{s}^{-1} \textrm{Mpc}^{-1}.
\end{equation}
The uncertainties in each estimate and in the final one are high
(also if reduced in this last one with respect to the single
systems), since it has to take into account all the degeneracies;
however,  adding other physical constraints and strengthening some
of those already used, it is possible to reduce the errors
ulteriorly. Also for the lens parameters there is this kind of
problem, above all in the orientation of the lens galaxy and the
strength parameters of some models. We think that this kind of
``marginalized'' estimate of $H_{0}$ over a large sample of models
can help us to overcome the problem of the lack of a correct
independent knowledge of the lens model.

\subsection{Application of the two-system model}
\begin{figure}
\centering
\resizebox{7cm}{!}{\includegraphics{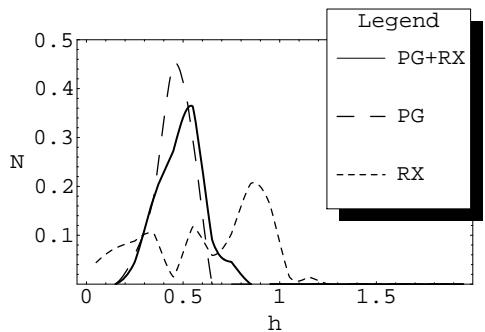}}
\caption{Distributions of all recovered values of the Hubble constant for the model withb $F(\theta)$ unknown, applied to the two systems.} \label{PGplusRX}
\end{figure}

In addition to the usual analysis done for each system, we apply
to PG 1115+080 and RX J0911+0551 the new method to simultaneously
fit the general elliptical model $\psi =F(\theta) r^{\alpha}$ with
external shear. In Table \ref{tab_common_h_PG_RX} we collect the
results, adding the values obtained for the single systems fitted
with the same potential. The uncertainties in the estimated value
for RX J0911+0551 are very high as well as for the other models
fitted in the paper, showing the presence of difficulties in the
fitting performed with these simple models. The angular trends for
the {\it two-system model} and the single systems are similar and
there is an agreement for the estimated $\alpha 's$.  The global
estimate of the Hubble constant is $H_{0}=49_{-11}^{+6} \ Km \
s^{-1} \ Mpc^{-1}$. Its determination is mainly determined by the
uncertainties in the estimate of $H_0$ for PG 1115+080, since the
distribution of the recovered for RX J0911+0551 is flat, giving us
a little quantity of information (see, for instance, Fig.
\ref{PGplusRX}).

\begin{table}
\begin{center}
\caption{Estimated parameters for the fitting of the elliptical potential with not fixed angular part ({\it two-systems model}), and comparison with the results obtained
fitting the same potential to the single
systems.}\label{tab_common_h_PG_RX} \vspace{0.5cm}
\begin{tabular}{|@{}c@{}|@{}c@{}|@{}c@{}|@{}c@{}|}
\hline
\rule{0pt}{2.4ex} $$ & {\it Two-systems} & \, {\it PG 1115+080} & \, {\it RX J0911+0551} \\
\hline \hline
\rule{0pt}{2.4ex} $r_{s}^{(1)}\hspace{0.08cm}('')$ & $0.25_{-0.14}^{+0.36}$  & $0.22_{-0.17}^{+0.12}$ & $-$\\
\hline
\rule{0pt}{2.4ex} $\theta_{s}^{(1)}\hspace{0.08cm}(^{o})$ & $23_{-38}^{+26}$ & $31_{-41}^{+26}$ & $-$\\
\hline
\rule{0pt}{2.4ex} $\gamma^{(1)}$ & $0.10_{-0.05}^{+0.01}$ & $0.09_{-0.06}^{+0.03}$ & $-$\\
\hline
\rule{0pt}{2.4ex} $\theta_{\gamma}^{(1)}\hspace{0.08cm}(^{o})$ & $-108_{-31}^{+9}$ & $-113_{-30}^{+8}$ & $-$\\
\hline
\rule{0pt}{2.4ex}  $\alpha^{(1)}$ & $1.13_{-0.15}^{+0.19}$ & $1.17_{-0.21}^{+0.22}$ & $-$\\
\hline
\rule{0pt}{2.4ex} $r_{s}^{(2)}\hspace{0.08cm}('')$ & $0.24_{-0.18}^{+0.12}$  & $-$ & $0.39_{-0.30}^{+0.28}$\\
\hline
\rule{0pt}{2.4ex} $\theta_{s}^{(2)}\hspace{0.08cm}(^{o})$ & $108_{-18}^{+15}$ & $-$ & $100_{-42}^{+22}$\\
\hline
\rule{0pt}{2.4ex} $\gamma^{(2)}$ & $0.11_{-0.10}^{+0.14}$ & $-$ & $0.25_{-0.16}^{+0.13}$\\
\hline
\rule{0pt}{2.4ex} $\theta_{\gamma}^{(2)}\hspace{0.08cm}(^{o})$ & $171_{-28}^{+24}$ & $-$ & $184_{-26}^{+18}$\\
\hline
\rule{0pt}{2.4ex}  $\alpha^{(2)}$ & $1.59_{-0.12}^{+0.13}$ & $-$ & $1.42_{-0.40}^{+0.35}$\\
\hline
\rule{0pt}{2.4ex} $h$ & $0.49_{-0.11}^{+0.06}$ & $0.45_{-0.11}^{+0.06}$ & $0.62_{-0.38}^{+0.28}$\\
\hline
\end{tabular}
\end{center}
\end{table}

These final values, but in particular the first one, are in good
agreement with the previous ones in the literature. The results
obtained by some of us in previous papers are also consistent with
the present ones: in CCRP01 it was obtained
\mbox{$H_{0}=56_{-11}^{+12} \hspace{0.1 cm} \textrm{km}
\hspace{0.05 cm} \textrm{s}^{-1} \textrm{Mpc}^{-1}$}, using an
elliptical potential without external shear and only PG 1115+080,
while in CCRP02 it has been found a value of
\mbox{$H_{0}=58_{-15}^{+17} \hspace{0.1 cm} \textrm{km}
\hspace{0.05 cm} \textrm{s}^{-1} \textrm{Mpc}^{-1}$}, fitting a
general elliptical potential to PG 1115+080 and B 1422+231. Using
the pixellated lens method, Williams \& Saha the results from PG
1115+080 and B 1608+656 are combined to finally get $H_{0}=61
{\pm} 11 \hspace{0.1 cm} \textrm{km} \hspace{0.05 cm}
\textrm{s}^{-1} \textrm{Mpc}^{-1}$, in good agreement with our
result. Using a $\chi^{2}$ minimization applied to B1608+656,
\cite{Koop-Fass} get $H_{0}=65_{-6}^{+7} \hspace{0.1 cm}
\textrm{km} \hspace{0.05 cm} \textrm{s}^{-1} \textrm{Mpc}^{-1}$,
only marginally in agreement with our result. In
\cite{Treu-Koopmans} it is obtained $H_{0}=59_{-7}^{+12}{\pm} 3
\hspace{0.1 cm} \textrm{km} \hspace{0.05 cm} \textrm{s}^{-1}
\textrm{Mpc}^{-1}$, modelling PG 1115+080 with two different
components so as to describe the luminous part and the  dark halo,
and using information by stellar dynamics.

\section{Conclusions}\label{conclusion}

In this paper we have presented a numerical method able to estimate lensing
parameters and Hubble constant for a wide class of models. The model
parameters as well as the Hubble constant have been estimated using as
constraints the image positions and the time delay ratios. We used two
classes of models: separable elliptical models and constant
mass\,-\,to\,-\,light ratio profiles, adding an external shear to take into
account the presence of an external group of galaxy. For these models we
solved the system composed by the combinations of lens equations and two
time delay ratios, selecting the solutions by means of suitable physical
constraints.  For each model and each parameter, we obtain an ensemble of
values: we used the mean as better estimate and a confidence level of
$68\%$ as error. In order to reduce the uncertainty due to the lens models
on the estimation of the Hubble constant, we {\it marginalized} over all
the models collecting the complete data set and obtained a final estimate
of $H_0$ and its error, which does not result dramatically underestimated,
in this way, because of an {\it a priori} choice of the model. To test the
code we created simulated systems, being able to recover the correct values
of parameters.

After the encouraging testing of the codes, we have then applied
them to two real systems for which  a measure of time delays has
been possible: PG 1115+080 and RX J0911+055. For PG 1115+080 it
was possible to get an estimate of $H_{0}= 58 {\pm} 27 \hspace{0.1
cm} \textrm{km} \hspace{0.1 cm}\textrm{s}^{-1} \textrm{Mpc}^{-1}$,
consistent with other results in the literature, obtained using
different techniques. For example, \cite{Courbin97,Ke-Ko97}  show
that the isothermal and pseudoisothermal models predict low values
of $H_{0}$, while the constant mass\,-\,to\,-\,light ratio ones
generate higher values. We can verify these results using Hubble
and de Vaucouleurs models, finding that a simple elliptical
isothermal model as {\it Model 2} predicts a very low value of
$H_{0}$, in agreement with the value $\sim 40 \hspace{0.1 cm}
\textrm{km} \hspace{0.1 cm}\textrm{s}^{-1} \textrm{Mpc}^{-1}$
obtained in  \cite{Impey-et-al98}. For RX J0911+0551, only the
elliptical profile allows to fit the image configuration. These
trends are also confirmed by the simulations.

As previously said, we ``marginalize'' over the models since we do
not know the ``correct'' form of the lens model, and hence we have
thought to overcome this difficulty in this way. The combination
of the two final distribution can help to reduce the uncertainties
and to obtain more information from more lens systems, thus should
avoid the problems in the fitting of the single systems. The
combined estimate is $H_{0}= 56 {\pm} 23 \hspace{0.1 cm}
\textrm{km} \hspace{0.1 cm}\textrm{s}^{-1} \textrm{Mpc}^{-1}$

The uncertainty in the final estimate can be further reduced
adding other models and including in the statistics other lensed
systems, consistently with the uncertainty obtained using other
methods (see, for instance, \cite{Wi-Saha2000}). If we consider
the contribution of the smoothness parameter $\tilde{\alpha}$, the
change in $H_{0}$ can be very high and comparable with uncertainty
in our estimates of $H_{0}$. For example, the variability for RX
J0911+0551 is $\sim 30-40\%$ (using these particular cosmological
models), with a comparable uncertainty in modelling.

The general method developed in this paper can be used to do more,
allowing to obtain an estimate of $H_0$ that is assumed for
hypothesis as the `same' for all the lensed systems (see
\cite{SW2004}). We can in fact fit simultaneously the two systems,
using a general elliptical potential with a not fixed angular part
and a shared $H_{0}$. The final estimate is $H_{0}=49_{-11}^{+6} \
\textrm{Km} \ \textrm{s}^{-1}\ \textrm{Mpc}^{-1}$, lower than the
result obtained by means of the marginalization of the 5 models
already analyzed in the paper, but in agreement within the
uncertainties. The hypothesis of a common $H_0$ is ambitious and
very strong, since we don't know if different lens system can be
fitted in the same manner by using the same lens model and the
same $H_0$. In fact, fitting different lens models, we have seen
that different ones for the two lensed systems give us different
values of $H_0$. But we think that a ``more-system'' model can
help to obtain a reasonable estimate.

Further improvements are possible. It will possible to use other
different models, for which it is not possible to write the
potential in a simple form, such as NFW profiles
(\cite{Navarro96,Navarro97}), or more complex ones (also realizing
more-system models). To take into account the different components
of the lensing galaxy, we can use different profiles to describe
their components; for example, it can be used a nearly isothermal
model to describe the dark halo and a de Vaucouleurs one to
describe the luminous profile. Then, we could also use an
exponential profile to account for a thin disk, that elliptical
galaxies sometimes seem to have. It is necessary a more accurate
modelling of RX J0911+0551 in order to give a better bound on the
estimated $H_0$ for this system, since we checked that it furnish
a little information and a little statistical weight. Then, it is
possible to shape double lensed system, also using the flux ratio
to better constrain the lens models to systems: we have to use a
function of merit similar to that used for the two-system model to
allow to have a necessary number of equations.

Finally, of course, the application of the single modelling (after
marginalization) and other two-systems ones to other lens systems with
measured time delays can allow to get a more and more accurate and precise
estimate of $H_{0}$, eliminating biases and errors linked to the use of
each lens model.

\section*{acknowledgements}

It is a pleasure to thank P. Scudellaro for the interesting
discussions we had during the making of this work and the referee
P. Saha for his significative suggestion in order to improve the
paper.

\end{document}